\documentclass{article}
\usepackage{amsmath}
\usepackage{amssymb}
\usepackage{geometry}
\usepackage{array}
\usepackage{CJK}
\usepackage{cite}
\usepackage{setspace}
\usepackage{extarrows}
\usepackage{amsthm}
\usepackage{fancyhdr}
\usepackage{enumerate}
\usepackage{bm}
\usepackage{color}
\usepackage{xcolor}
\usepackage{eso-pic}
\usepackage{mathrsfs}
\usepackage{braket}
\usepackage{graphicx}
\usepackage{float}
\usepackage{subfigure}
\usepackage{amssymb, amsmath,amsfonts}
\usepackage{graphicx}
\usepackage{xcolor}
\usepackage{color}
\usepackage[pdftex, bookmarks=true,colorlinks,linkcolor=red,urlcolor=blue,citecolor=blue]{hyperref}
\usepackage[normalem]{ulem}
\usepackage{cite}
\usepackage{subcaption}
\usepackage{slashed}

\renewcommand{\thesubfigure}{(\roman{subfigure})}
\makeatletter \renewcommand{\@thesubfigure}{\thesubfigure \space}
\renewcommand{\p@subfigure}{} \makeatother
\makeatletter

\newcommand{\Rmnum}[1]{\expandafter\@slowromancap\romannumeral #1@}
\makeatother
\usepackage{cite}
\newcommand{\enabstractname}{Abstract}

\usepackage[numbers,sort&compress]{natbib}
\newcommand{\upcite}[1]{\textsuperscript{\textsuperscript{\cite{#1}}}}
\usepackage{setspace}

\UseRawInputEncoding
\pdfoutput=1 
\usepackage{amssymb, amsmath,amsfonts}
\usepackage{graphicx}
\usepackage{xcolor}
\usepackage{color}
\usepackage[pdftex, bookmarks=true,colorlinks,linkcolor=red,urlcolor=blue,citecolor=blue]{hyperref}
\usepackage[normalem]{ulem}
\usepackage{cite}
\usepackage{subcaption}
\allowdisplaybreaks
\usepackage{enumerate}
\usepackage[normalem]{ulem}

\makeatletter\@addtoreset{equation}{section}\makeatother

\newcommand{\preprint}[1]{\begin{table}[t]  
             \begin{flushright}               
             {#1}                             
             \end{flushright}                 
             \end{table}}                     
\renewcommand{\title}[1]{\vbox{\center\LARGE{#1}}\vspace{5mm}}
\renewcommand{\author}[1]{\vbox{\center#1}\vspace{5mm}}

\newcommand{\address}[1]{\vbox{\center\em#1}}

\usepackage{bm}

\def\be{\begin{eqnarray}}
\def\ee{\end{eqnarray}}
\def\bea{\begin{eqnarray}}
\def\eea{\end{eqnarray}}

\def\Dslash{\,\,{\raise.15ex\hbox{/}\mkern-12mu D}}
\def\Dbarslash{\,\,{\raise.15ex\hbox{/}\mkern-12mu {\bar D}}}
\def\delslash{\,\,{\raise.15ex\hbox{/}\mkern-9mu \partial}}
\def\delbarslash{\,\,{\raise.15ex\hbox{/}\mkern-9mu {\bar\partial}}}
\def\pslash{\,\,{\raise.15ex\hbox{/}\mkern-9mu p}}
\def\calDslash{\,\,{\raise.15ex\hbox{/}\mkern-12mu {\cal D}}}

\def\lae{\mathrel{\mathop{\smash{\lower .5 ex \hbox{$\stackrel<\sim$}}}}}
\def\lae{\mathrel{\mathop{\smash{\lower .5 ex \hbox{$\stackrel>\sim$}}}}}

\begin{document}
\unitlength = .8mm

\begin{titlepage}
\vspace{.5cm}
\preprint{}
\begin{center}
\hfill \\
\hfill \\
\vskip 1cm\

\title{\bf Induced magneto-conductivity in a two-node Weyl semimetal under Gaussian random disorder}
\vskip 0.5cm

{Chuan-Xiong Xu$^{a}$},{Hao-Ping Yu$^{b}$},{Mei Zhou$^{a}$},{Xuanting Ji$^{a,c}$}\footnote{Email: {\tt jixuanting@cau.edu.cn}}

\address{${}^a$Department of Applied Physics, College of Science, \\
China Agricultural University, Beijing 100083, China}
\vspace{-10pt}

\address{${}^b$Department of Applied Mechanics, College of Science, \\
China Agricultural University, Beijing 100083, China}
\vspace{-10pt}

\address{${}^c$School of Physical Sciences, and CAS Center for Excellence in Topological Quantum Computation, University of Chinese Academy of Sciences, 
Beijing 100049, China}
\vspace{-10pt}

\end{center}
\vspace{-10pt}
\abstract{Measuring the magnetoconductivity induced from impurities may help determine the impurity distribution and reveal the structure of a Weyl semimetal sample. To verify this, we utilized the Gaussian random disorder to simulate charged impurities in a two-node Weyl semimetal model and investigate the impact of charged impurities on magnetoconductivity in Weyl semimetals. We first compute the longitudinal magnetic conductivity and find that it is positive and increases proportionally with the parameter governing the Gaussian distribution of charged impurities, suggesting the presence of negative longitudinal magnetoresistivity (NLMR). Then we consider both the intra-valley and inter-valley scattering processes to calculate the induced transverse magnetoconductivity in the model. Our findings indicate that both inter-valley and intra-valley scattering processes play important roles in calculating the transverse magnetoconductivity. The locations of Weyl nodes can also be determined by magnetoconductivity measurements. This is possible if the magnetic field strength and the density of charged impurities are known. Alternatively, the measurement of magnetic conductivity may reveal the distribution of charged impurites in a given sample once the locations of the Weyl nodes have been determined. These findings can aid in detecting the structure of a Weyl semimetal sample, enhancing comprehension of magnetotransport in Weyl semimetals, and promoting the development of valley electronics.}

\vfill

\end{titlepage}

\begingroup
\hypersetup{linkcolor=black}
\tableofcontents
\endgroup

\section{Introduction}
\par
Weyl fermions were first introduced by Hermann Weyl in 1928. These fermions are characterized by their chirality and lack of mass. Up to now, no candidates have been found within the standard model of particle. However, it has been discovered that low energy excitations in certain condensed matter systems, known as Weyl semimetals, can realize Weyl fermions. The initial theoretical proposal for the Weyl semiemtal occurred in 2011\upcite{PhysRevB.83.205101}. Subsequently, in 2015, the first authentic Weyl semimetals were achieved within the $TaAs$ family \upcite{weng_weyl_2015,lv_experimental_2015,sun_topological_2015,yang_weyl_2015,levy_optical_2020}. These findings encourage the search for additional candidates of the Weyl semimetal. Other materials, such as $YbMnBi_2$(Type-I)\upcite{borisenko_time-reversal_2019}, $Mo_xW_{1-x}Te_2$(Type-II)\upcite{soluyanov_type-ii_2015,Deng_2016,PhysRevLett.117.266804}, and $(TaSe_4)_2I$(Type-III)\upcite{PhysRevB.103.L081402}, have been demonstrated as viable candidates for Weyl semimetals.

Weyl fermions possess chirality and always appear in pairs. The Weyl fermions in the Weyl semimetal can manifest the negative longitudinal magnetic resistance (NLMR)\upcite{burkov_chiral_2014,PhysRevB.89.085126,PhysRevB.88.104412}. This is done by applying an electric field parallel to the direction of a magnetic field in the Weyl semimetal, the Weyl fermions of one specific chirality are pumped to their partner of opposite chirality, according to the no-go theorem.\upcite{NIELSEN1983389} The NLMR, an anomalous transport due to chiral anomalies, which have been observed in several experiments\upcite{zhang_signatures_2016,WOS:000812345900001,kawasuso_robustness_2023-1,Ahmad_2021,Zhang_2021,Wadge_2022,zhang_observation_2015,takiguchi_quantum_2020,cohn_magnetoresistance_2020,ong_experimental_2021,kundu_magnetotransport_2020,labarre_evidence_2020,shekhar_extremely_2015}.
However, positive and negative longitudinal magnetoresistance were observed in some experiments\upcite{zhang_signatures_2016}, indicating that magnetoconductivity measurements are sample-dependent, especially disorder-dependent\upcite{takiguchi_quantum_2020}. Additionally, in weak magnetic fields, some experiments have shown a peak-like nonparabolic negative magnetoresistance\upcite{cohn_magnetoresistance_2020}. Theoretical explanations are needed for these experimental phenomena.

Theoretically, conductivity can be viewed as the collision process between impurity electrons and system electrons near the Fermi energy level. For the undoped case of Weyl semimetal, the Fermi energy level is situated at the Weyl nodes\upcite{PhysRevB.83.205101,PhysRevLett.107.127205,PhysRevLett.119.036601}. The charged impurities introduce additional charges to the energy bands, causing a shift in the Fermi level away from the Weyl nodes. Thus, the Fermi energy levels exhibit extensive variation among the diverse samples evaluated in experiments and are highly responsive to alterations in sample properties. To accurately reproduce and predict experimental results in theoretical calculations, one must consider the diverse distributions of impurities present. Previously, there have been a large number of theoretical studies\upcite{jiang_chirality-dependent_2021,PhysRevB.92.174202,PhysRevB.92.205113,PhysRevB.85.241101,PhysRevB.88.104412,PhysRevB.92.045203,PhysRevB.97.064203,yan_topological_2017,lu_weak_2015,zhang_double-weyl_2018,zhang_linear_2016,sbierski_quantum_2014,yokouchi_giant_2023,yu_editorial_2023}, several studies have examined the impact of distant disorder on the density of states in a weak magnetic field\upcite{PhysRevB.92.174202}, as well as the density of the states, Landau level broadening, and magneto-resistivity in a transverse magnetic field with short-range or charged impurities\upcite{PhysRevB.92.205113}. However, most of these studies are based on the one-node model, which assumes the existence of only a single Weyl fermion\upcite{PhysRevB.92.174202,PhysRevB.92.205113,PhysRevB.92.045203,lu_weak_2015,sbierski_quantum_2014}. A one-node model cannot present the chiral anomaly since the chiral anomaly requires at least one pair of Weyl fermions. Therefore, studying NLMR in one-node models is not enough.

The two-node model for a Weyl semimetal is the simplest extension of the one-node model. It consists of two Weyl fermions with opposite chirality, allowing for the study of the chiral anomaly. Additionally, two-node model naturally introduces the valley degree of freedom of the Weyl fermion, allowing for the simultaneous study of inter-valley and intra-valley scattering processes. As pointed in\upcite{ji_competition_2017,ji_effect_2018,shao_magneto-conductivity_2019,feng_magneto-conductivity_2020}, the inter-valley scattering processes contribute significantly to the magnetoconductivity in the presence of an impurity distribution that obeys a screened Coulomb potential or a Dirac $δ$ potential. However, the screened Coulomb disorder and the Dirac $δ$ potential may not accurately replicate the charged impurities in practical scenarios. Therefore, we propose that the impurities follow a Gaussian distribution. This distribution may provide a more accurate representation of the distribution of charged impurities and model a more realistic Weyl semimetal sample, which is used in experiments. A random Gaussian potential was used to simulate charged impurities has been studied in\upcite{zhang_linear_2016}. In \upcite{zhang_linear_2016}, the authors investigated the longitudinal magnetoresistance in a two-node model Weyl semimetal. However, they did not provide details on transverse magneto-conductivity or address inter-valley scattering.

With all of the backgrounds mentioned above, in this work we utilized random Gaussian potential to model the distribution of charged impurities in a two-node Weyl semimetal model. Taking into account the contributions from inter-valley scattering processes, we studied both longitudinal and transverse magnetoconductivity. Based on the calculation of the longitudinal conductivity, it is evident that the conductivity is positive definite and increases with magnetic flux density, indicating the presence of negative longitudinal magnetoresistivity, supporting the theoretical prediction of chiral anomalies. The results of the transverse magnetoconductivity show that the inter-valley scattering process has non negligible contributions to the magnetoconductivity.  Furthermore, we carefully analyzed the impact of the length scale of the Gaussian distribution, magnetic field strength, and the position of Weyl nodes on transverse magnetoconductivity. We find that the transverse magneto-conductivity
is inversely proportional to the magnetic flux density. Besides, we were able to replicate the peaks observed in \upcite{cohn_magnetoresistance_2020} by adjusting the magnetic flux density and the length scale of the Gaussian distribution in our model. Our results indicate that the peaks are located near the Weyl nodes, so it is possible to infer the structure of a Weyl semimetal sample just by measuring its magnetoconductivity, which may be verified in future experiments.

This paper is organized as follows: In section \ref{sec2}, we review a two-node model of the Weyl semimetal and present the eigen-energy as well as the eigen-states. In section \ref{sec30}, we interpret a mechanism for shifting Fermi energies. Then we derive the formulas and show the numerical results for the longitudinal and transverse magnetoconductivity in \ref{sec31} and \ref{sec32}. Possible applications of magnetic conductivity measurement are discussed in \ref{sec33}. Section \ref{sec4} is devoted to conclusions and discussions.

\section{A two-node model of the Weyl semimetal}\label{sec2}
In this section we will review a two-nodes model for the Weyl semimetal raised in\upcite{ji_competition_2017,ji_effect_2018}, in the absence of an external magnetic field, the Hamiltonian of the two-nodes model can be written as
\begin{equation}
\mathcal{H}=\nu\hbar{v_F}[(\boldsymbol{k}+\nu{\boldsymbol{k_c}})\cdot{\boldsymbol{\sigma}}].
\label{eq1}
\end{equation}
The energy spectrum of eq.\ref{eq1} spreads in momentum space with two valleys corresponding to two Weyl nodes with opposite chiralities, we use $\nu$ to indicate this with $\nu=1$ for $K$ valley and $\nu=-1$ for $K^{'}$ valley. $\hbar$ is reduce planck constant, $v_F$ is Fermi velocity. $\boldsymbol{k_c}=(0,0,±k_c)$ in momentum space represent the positions of the two nodes. In the following we will write eq.\ref{eq1} into two parts, each part corresponds to one valley:
\begin{equation}
{\tilde{H}}_K={\mathcal{H}_K}-\hbar{v_F}{\boldsymbol{k_c}}\cdot{\boldsymbol{\sigma}},
\label{eq2}
\end{equation}
\begin{equation}
{\tilde{H}}_{K^{'}}={\mathcal{H}_{K^{'}}}-\hbar{v_F}{\boldsymbol{k_c}}\cdot{\boldsymbol{\sigma}}.
\label{eq3}
\end{equation}
These two Hamiltonians allow for a more intuitive study of intra-valley and inter-valley scattering processes. We will give the eigenstates and eigenenergies of them in the following.

For \textit{K} valley, the eigenenergies are
\begin{equation}
	{\tilde{E}}_{±K}=±\hbar{v_F}\sqrt{{k_x}^2+{k_y}^2+{k_z}^2},
 \label{eq4}
\end{equation}
with eigenstates as:
\begin{equation}
\Psi_{+K}=\begin{pmatrix}
cos\frac{\theta}{2}\\
sin\frac{\theta}{2}e^{i\varphi}
\end{pmatrix},
\Psi_{-K}=\begin{pmatrix}
sin\frac{\theta}{2}\\
-cos\frac{\theta}{2}e^{i\varphi}
\end{pmatrix},
\label{eq5}
\end{equation}
where $\pm$ represent the conduction and valence bands, respectively.
$\theta$ and $\varphi$ are determined as:
\begin{equation}
cos{\theta}=\frac{k_z}{\tilde{|E|}},tan{\varphi}=\frac{k_x}{k_y}.
\label{eq6}
\end{equation}
\par
When a magnetic field $\textbf{B}= (0, 0, B)$ is applied along the $z$-direction, Landau bands form and disperse with $k_z$. Here we choose the Landau gauge $\textbf{A} = (yB, 0, 0)$ and we have:
\begin{equation}
\boldsymbol{k}=(k_x-\frac{eB}{\hbar}y,-i{\partial_y},k_z).
\label{eq7}
\end{equation}
To obatin the eigenstates, we need to rewrite the Hamiltonian with the ladder operators:
\begin{equation}
\textit{a}=-[(y-{\textit{l}_B}^{2}k_x)/{\textit{l}_B}+\textit{l}_B{\partial_y}]/\sqrt{2},
\label{eq8}
\end{equation}
\begin{equation}
\textit{a}^{\dagger}=-[(y-{\textit{l}_B}^{2}k_x)/{\textit{l}_B}-\textit{l}_B{\partial_y}]/\sqrt{2},
\label{eq9}
\end{equation}
where $\textit{l}_B=\sqrt{\hbar/eB}$ is the magnetic length. Then Eq.\ref{eq2} becomes
\begin{equation}
\tilde{H}_K=\hbar{v_F}\begin{pmatrix}
\textit{k}_z&-\frac{\sqrt{2}}{\textit{l}_B}\textit{a}\\
-\frac{\sqrt{2}}{\textit{l}_B}\textit{a}^{\dagger}&-{\textit{k}_z}
\end{pmatrix}.
\label{eq10}
\end{equation}
Then the eigenenergies for the original Hamiltonian are
\begin{equation}
	{E}_{{\pm}K}^n=\tilde{E}_{{\pm}K}^n+\hbar{v_F}k_c,
 \label{eq11}
\end{equation}
\begin{equation}
	\tilde{E}_{{\pm}K}^n=\pm\hbar{v_F}\sqrt{{k_z^2}+2n/{\textit{l}_B}^2}.
 \label{eq12}
\end{equation}
Here $n=0,1,2\dots$ is the eigenvalue of the operator $\hat{N}=a^{\dagger}a$, corresponding to different Landau levels.

Similarly, the corresponding eigenenergies for $K^{'}$ valley are obtained
\begin{equation}	{E}_{{\pm}K^{'}}^n=\tilde{E}_{{\pm}K^{'}}^n+\hbar{v_F}k_c.
 \label{eq13}
\end{equation}
\begin{equation}
\tilde{E}_{{\pm}K^{'}}^n=\mp\hbar{v_F}\sqrt{{k_z^2}+2n/{\textit{l}_B}^2},
 \label{eq14}
\end{equation}
More details of calculations can be found in the supplementary material.
\section{Formalism and results}
\label{sec3}
In this section, we presents our main results. In subsection \ref{sec30} we show how the Fermi energy moves away from the Weyl nodes by adding impurities. We then show the numerical results for longitudinal and transverse
and transverse magnetoconductivity in subsections \ref{sec31} and \ref{sec32}. Our main results are presented in subsection \ref{sec33}, where we explain how to determine the structure of a Weyl semimetal sample by measuring its magnetic conductivity.
\subsection{The shift of the Fermi energy}
\label{sec30}
In an undoped Weyl semimetal, the Fermi level is located precisely at the Weyl nodes. The addition of donor and acceptor impurities introduces extra charges into the energy bands, shifting the Fermi level away from the Weyl nodes. To illustrate this physical scenario, we define $n_d$ and $n_a$ as the densities of the donor and acceptor, respectively, to calculate the density of excess electrons. With the definition \upcite{ji_effect_2018}:
 \begin{equation}
 	n_{imp}(\mu,T)=n_d-n_a,
  \label{eq15}
 \end{equation}
the excess electron density is calculated via
\begin{equation}
 n_{imp}(\mu,T)=g\sum_{\nu=\pm}\sum_{k_z}[f_0(\nu\hbar{v_F}k_z-\mu)-f_0(\nu\hbar{v_F}k_z)]
  \label{eq16}
 \end{equation}
Here $g=2\times{g_0}$ represents the spin and Landau level degeneracy, where $g_0=eB/2\pi\hbar$ is the the degeneracy of every per unit area. By combine the Fermi-Dirac distribution $f_0(x)=\frac{1}{e^{x}/{k_B}T+1}$ we obtain the excess electron density as:
\begin{equation}
	n_{imp}(\mu,T)=2g\frac{{k_B}T\bm{ln}({\frac{e^{\mu/k_BT}+1}{2})}}{2}
 \label{eq17}
\end{equation}
In experiments, the value of $n_{imp}$ remains constant for each sample containing both donors and acceptors. Hence, by taking $\mu=E_F$, we obtain the new location of the Fermi energy as\footnote{In this paper, we did not consider the effect of temperature. Therefore, in the following calculations, we will fix $T=3K$ without loss of generality.}
\begin{equation}
	E_F\equiv{k_BT\bm{ln}(2e^{\frac{\pi{n_{imp}}\hbar{v_F}}{2gk_BT}-1})}
 \label{eq18}
\end{equation}

\subsection{Longitudinal magneto-conductivity}
\label{sec31}
As the Fermi level leaves the Weyl nodes but does not reach the first Landau level, the transport is essentially involved with the $0$-th Landau band. The semiclassical conductivity of this case can be simply figured out by using the Einstein relation: $\sigma_{zz}=e^2\textit{N}_F\textit{D}$, where $\textit{N}_F=(1/2\pi{\textit{l}_B^2})(2/\pi\hbar{v_F})$, $\textit{D}=v_F^2\tau_{k_F}^{0,tr}$ is the diffusion coefficient. $\tau_{k_F}^{0,tr}$ is the transport time with $k_F$ is the Fermi wave number. According to semi-classical Boltzmann transport theory, under the first-order Born approximation, for the scattering among the states on the Fermi surface of the $0$-th band, we obtain the relaxation time by
\begin{equation}
\frac{\hbar}{\tau_{k_z}^{0,tr}}=2\pi\sum\limits_{k_x^{'}k_z^{'}}\bra{}U_{k_x,k_z,{k_x^{'}},{k_z^{'}}}^{0,0}\ket{ ^2}\Lambda\delta(E_F-{E_{k_z^{'}}^0})\times(1-\textit{v}_{0,k_z^{'}}^z/\textit{v}_F),
\label{eq19}
\end{equation}
where $\textit{E}_F$ is the Fermi energy, and $\textit{E}_{k_z^{'}}^0$ is the eigenenergy of the $0$-th Landau level. $U_{k_x,k_z,{k_x^{'}},{k_z^{'}}}^{0,0}$ represent the scattering matrix, it can take place in the scattering process between the eigen state $\ket{0,k_x,k_z}$ and eigen state $\ket{0,k_x^{'},k_z^{'}}$. $\bra{}U_{k_x,k_z,{k_x^{'}},{k_z^{'}}}^{0,0}\ket{ ^2}$ in here is the corresponding scattering matrix elements. $\Lambda$ is introduced to correct the unphysical van Hove singularity near the band edge, which always take as $1$\upcite{ji_competition_2017}. ${v}_{0,k_z^{'}}^z$ is the velocity of an electron along the $z$-direction with the wave number $k_z^{'}$ in the $0$-th Landau band. For $z$-axis magnetic field, we note that only the diagonal elements of the velocity matrix with respect to the $z$-direction contribute to the longitudinal magneto-conductivity. On the other hand, only the off-diagonal elements of the velocity matrix with respect to the $x$-direction contribute to the transverse magneto-conductivity.
The longitudinal conductivity $\sigma_{zz}$ can be found from the formula
\begin{equation}
\sigma_{zz}=\frac{e^2\hbar}{2{\pi}V}\sum\limits_{k_x,k_z,i}Tr(\textbf{v}_{0,k_z}^{z}\textbf{G}_{0,k_z}^R\tilde{\textbf{v}}_{0,k_z}^{z}\textbf{G}_{0,k_z}^A),
\label{eq20}
\end{equation}
both $\textbf{v}$ and $\textbf{G}$ are diagnoal matrixes in the valley space, the subscript $0$ represents the $n=0$ Landau subspace.

We defined the ${v}_{0,k_z}^{z\textbf{K}}$ as the ${z}$-direction velocity of an electron in the $0$-th band, and $\tilde{v}_{0,k_z}^{z\textbf{K}}$ is the dressed velocity after taking into account the vertex correction. Here we have
\begin{equation}
{v}_{0,k_z}^{z\textit{K}}=\partial{E}_{k_z}^{0\textit{K}}/\hbar\partial{k_z}=-{v_F}.
\label{eq21}
\end{equation}
In this case, we fixed the Fermi Energy at the $0$-th Landau band, and all the Landau bands in the $K^{'}$ cone are fully occupied. Hence, we have
\begin{equation}
\sigma_{zz}=\frac{e^2\hbar}{2{\pi}V}\sum\limits_{k_x,k_z}({v}_{0,k_z}^{z}\tilde{{v}}_{0,k_z}^{z}{G}_{0,k_z}^R{G}_{0,k_z}^A),
\label{eq22}
\end{equation}
The retarded/advanced Green's function for the $0$-th band are $G_{0\textbf{K}}^{R/A}=1/(E_F-E_{k_z}^{0\textit{K}}\pm{i\hbar/2\tau})$ where $\tau$ is the corresponding momentum relaxation time which is related to the transition probability(induced by the impurites) in \textit{K} valley at $n=0$ band. And $V=L_xL_yL_z$ here is the volume of a system.

In the diffusive regime, one can replace $G_{0\textit{K}}^R\tilde{v}_{0,k_z}^{z\textit{K}}G_{0\textit{K}}^A$ by $(2\pi/\hbar)\tau_{k_z}^{0\textit{K},tr}{v}_{0,k_z}^{z\textit{K}}\Lambda\delta(E_F-{E_{k_z}^{0\textit{K}}})$,
then we get
\begin{equation}
\sigma_{zz}=\frac{e^2}{V}\sum\limits_{k_z,k_x}({v}_{0,k_z}^{z\textit{K}})^2\tau_{k_z}^{0\textit{K},tr}\Lambda\delta(E_F-{E_{k_z}^{0\textit{K}}}).
\label{eq23}
\end{equation}

To calculate the transport time $\tau_{k_z}^{0\textit{K},tr}$, we need to do the Fourier transform of the Gaussian random potential $f(\textbf{r})=\textit{C}e^{-r^2/2d^2}$, which can be written as:
\begin{equation}
U(\textbf{q})=u_0e^{\frac{q^2d^2}{2}},
\label{eq26}
\end{equation}
where $u_0$ measures the strength, $d$ is the length scale of the Gaussian distribution parameter. After calculate the transport time $\tau_{k_F}^{0\textit{K},tr}$ at the Fermi surface, we obtain the final expression for $\sigma_{zz}$ as:
\begin{equation}
\sigma_{zz}=\frac{2e^2}{\textit{h}}\frac{(\hbar{v_F})^2(d^2+{\textit{l}_B}^2)}{{\textit{l}_B}^2{V_{imp}}e^{-(2k_F-k_c)^2d^2}},
\label{eq27}
\end{equation}
where $V_{imp}={n_{imp}}{u_0}^2$, $n_{imp}$ represent the doping degree. The calculation details can be found in the supplementary material.
\begin{figure}[h]
	\centering
	\includegraphics[width=0.8\textwidth]{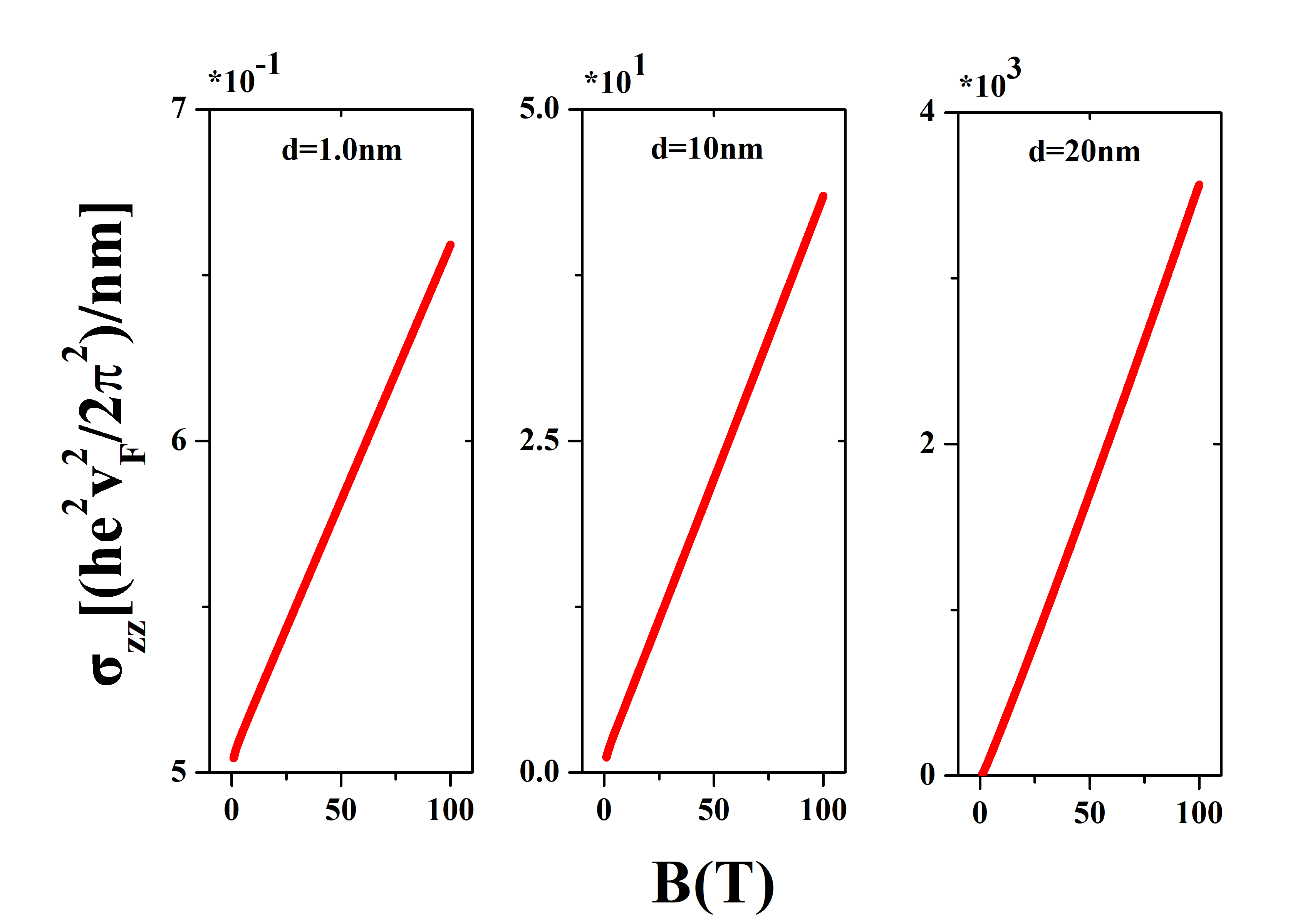}
	\caption{\small{The longitudinal magneto-conductivity versus the magnetic flux density for different $d$ value. Here $v_F=10^6m/s$, $k_c=0.1nm^{-1}$, $n_{imp}=1.0\times10^{15}cm^{-3}$.}}
	\label{fig1}
\end{figure}
\par
In Fig.\ref{fig1}, we present the variation of longitudinal magnetoresistance in the two-node model with different Gaussian distribution length scales, described by the parameter $d$, under the influence of random Gaussian potential doping at low temperature ($T=3K$, unless otherwise specified).  Consistent with previous studies, we obtained the positive longitudinal magnetoconductivity (negative longitudinal magnetoresistance). Here, we explore the influence of doping on $\sigma_{zz}$ by varying the parameter $d$ in the impurity potential. The results show that the variation of $\sigma_{zz}$ with magnetic induction intensity $B$ is linear, consist with the results in\upcite{ji_competition_2017, ji_effect_2018,shao_magneto-conductivity_2019, zhang_linear_2016}.

We further investigate the effect of the length scale of the Gaussian distributions by fixing the value of $d$ at $1.0nm$, $10nm$, and $20nm$, respectively. We find that the longitudinal magnetoresistance increases with increasing $d$, for $d$ in the range of $0.1nm$ to $10nm$ the $\sigma_{zz}$ increases slowly, while for $d$ greater than $10nm$ the $\sigma_{zz}$ increases faster. This phenomenon can be explained by the Heisenberg uncertainty principle, which states that larger impurities with a larger $d$ have a wider distribution, resulting in a smaller change in position ($\Delta x$) and a larger change in momentum ($\Delta p$). This leads to more intense collision processes and a larger longitudinal magneto-resistance (or smaller longitudinal magneto-conductivity).
\subsection{Transverse magneto-conductivity}
\label{sec32}
Similarly, we study in this subsection the transverse magneto-conductivity($\sigma_{xx}$). For consistency, the Fermi energy is assumed to be near the Weyl nodes, and the conductivity along the ${x}$-direction can be written as
\begin{equation}
	\sigma_{xx}=\frac{e^2\hbar}{2\pi{v_F}}\sum\limits_{\textbf{k}}Tr(\textbf{G}^R\textbf{v}_{\textbf{k}}^x\textbf{G}^A\textbf{v}_{\textbf{k}}^x).
 \label{eq28}
\end{equation}
As in the previous section, both $\textbf{G}$ and $\textbf{v}$ are matrixes, the derivation of Eq.\ref{eq28} is given in supplementary material, we have
\begin{equation}
	\sigma_{xx}=\frac{e^2\hbar}{2\pi{v_F}}\sum\limits_{k_x,k_z,i,i^{'}}({G_{0i}^R}{v_{0i,i+i^{'}}^x}{G_{1+i^{'}}^A}{v_{1+i^{'},0i}^x}).
 \label{eq29}
\end{equation}
After taking both the intra-valley and inter-valley scattering into account, we get the total transverse magneto-conductivity:
\begin{equation}
	\sigma_{xx}=\sigma_{xx,inter}^{\textit{K}}+\sigma_{xx,intra}^{\textit{K}}+\sigma_{xx,inter}^{\textit{K}^{'}}+\sigma_{xx,intra}^{\textit{K}^{'}}.
 \label{eq30}
\end{equation}
The relaxation time$(\tau)$ corresponding to different scattering processes(two kinds of intra-valley scattering and two kinds of inter-valley scattering) can be listed in supplementary material, and we can get the $\sigma_{xx}$ by
\begin{equation}
	\begin{aligned}
		&\sigma_{xx,intra}^{K}=\frac{e^2}{4\pi^2h}\frac{V_{imp}e^{-d^2(2k_F-k_c)^2}}{(d^2+l_B^2/2)^2}\frac{F_2^K(k_z)}{4}\\
		&\sigma_{xx,inter}^{K}=\frac{e^2}{4\pi^2h}\frac{V_{imp}e^{-d^2{k_c}^2}}{(d^2+l_B^2/2)}{F_1^K(k_z)}\\
		&\sigma_{xx,intra}^{K^{'}}=\frac{e^2}{4\pi^2h}\frac{V_{imp}e^{-d^2(2k_F-k_c)^2}}{(d^2+l_B^2/2)}{F_1^{K^{'}}(k_z)}\\
		&\sigma_{xx,inter}^{K^{'}}=\frac{e^2}{4\pi^2h}\frac{V_{imp}e^{-d^2{k_c}^2}}{(d^2+l_B^2/2)^2}\frac{F_2^{K^{'}}(k_z)}{4},
	\end{aligned}
 \label{eq31}
\end{equation}
we call $\textit{F}$s form factors, and derive them in supplementary material.
\par
\begin{figure}[h]
	\centering
	\includegraphics[width=0.95\textwidth]{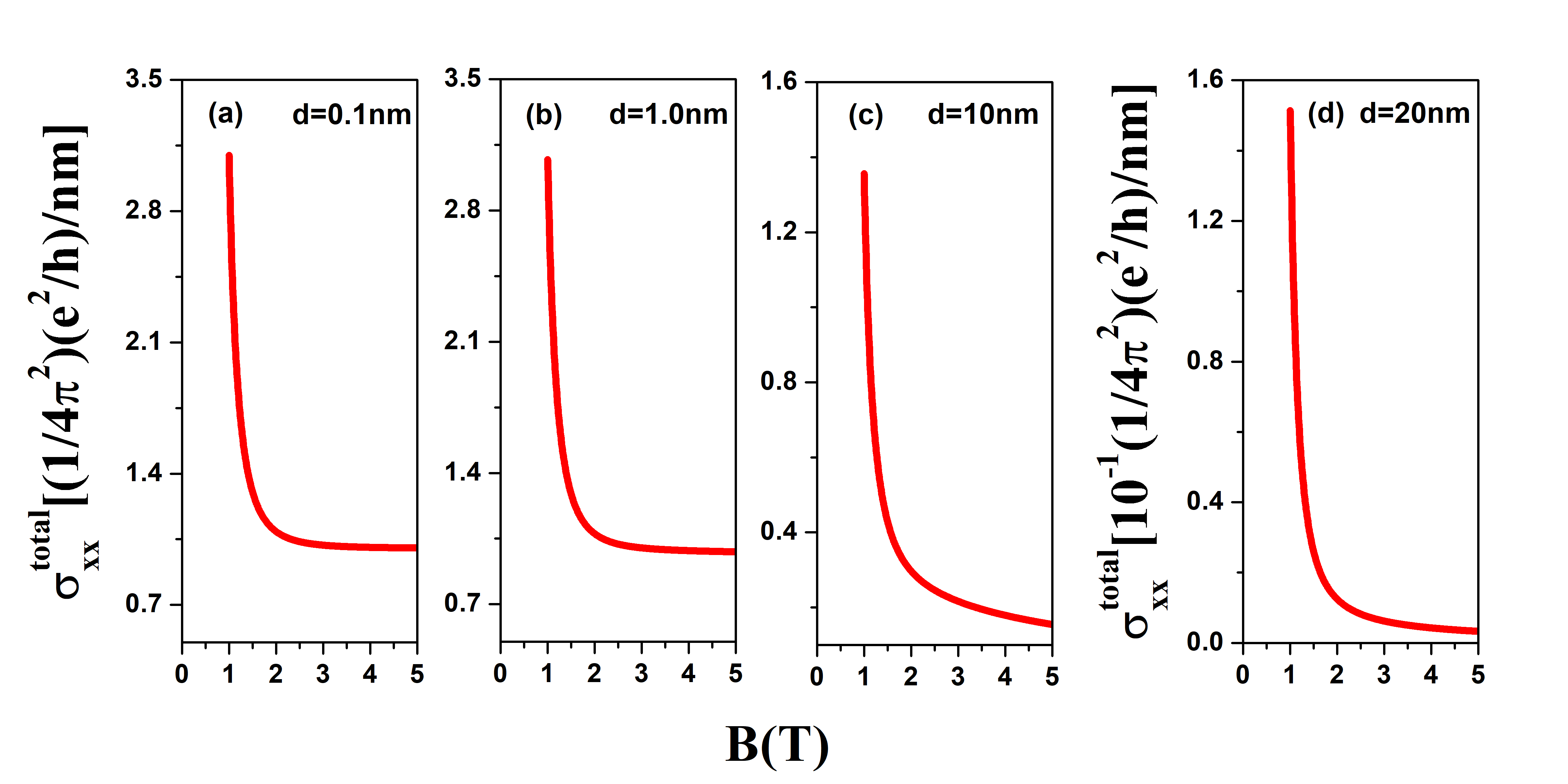}
	\caption{\small{The transverse magneto-conductivity, $\sigma_{xx}$ , as a function of $B$ at different $d$-value for (a) $d=0.1nm$, (b) $d=1.0nm$, (c) $d=10nm$, and (d) $d=20nm$, respectively. Here $k_c=0.1nm^{-1}$, $n_{imp}=1.0\times10^{15}cm^{-3}$.}}
	\label{fig2}
\end{figure}
In Fig.\ref{fig2}, we show the variation of the transverse magnetoconductivity as a function of the magnetic induction intensity $B$ for different Gaussian potentials (different parameter $d$). We find that the important change characteristics of the transverse magneto-conductivity are different from those of the longitudinal magneto-conductivity, and the transverse magneto-conductivity decreases with the increase of $d$, which is just the opposite of the longitudinal magneto-conductivity that will be inhibited with the increase of $d$. Fig.\ref{fig2}(a) and Fig.\ref{fig2}(b) demonstrate a minor change in transverse magneto-conductivity when the parameter $d$ is small, whereas Fig.\ref{fig2}(c) and Fig.\ref{fig2}(d) exhibit a significant change in transverse magneto-conductivity, This is consistent with results in ref\upcite{zhang_linear_2016}. Another important feature is clearly shown in the image, under the lower intensity of the magnetic field $(0-2T)$, the decline trend of the curve is steeper. The appearance of the  threshold can be understood as follows: the $l_B=\sqrt{\hbar/eB}\approx25.6/\sqrt{B} \bold{nm}$ corresponds to the Landau level degeneracy. When the magnetic field is weak, $l_B$ is large enough to be comparable with the scale of impurity distribution, resulting in more collisions between electrons and charged impurities. When the magnetic field flux exceeds 2T, the value of $l_B$ rapidly decreases, causing the electrons to become highly degenerate in a small region. Thus, collisions between electrons and impurities become localized, leading to a slowly change in transverse conductivity.
\begin{figure}[h]
	\centering
	\includegraphics[width=0.8\textwidth]{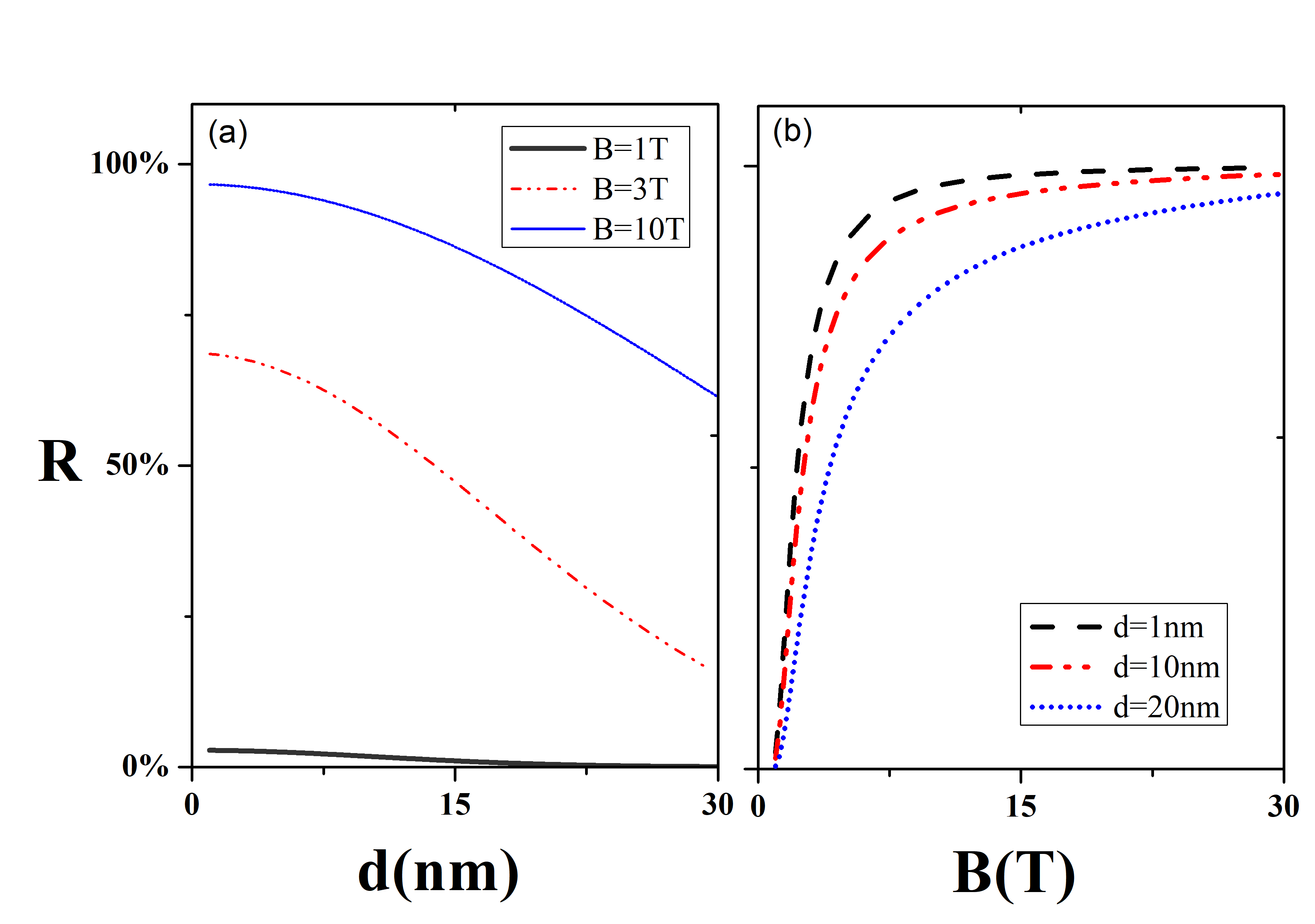}
	\caption{\small{The trend of ratio $R$ with $B$ and $d$ values are studied. Here $R=\frac{\sigma_{xx,inter}}{\sigma_{xx,intra}}$. In (a), the black thin line is for $B=1T$; the red dotted line is for $B=3T$, and the blue thick line is for $B=10T$, respectively. In (a), the black thin line is for $B=1T$; the red dotted line is for $B=3T$, and the blue thick line is for $B=10T$, respectively. In (b), the black dashed line represents $d=1.0nm$, the red dash-dot line represents $d=10nm$, and the blue dotted line represent $d=20nm$. Here $k_c=0.1nm^{-1}$, $n_{imp}=1.0\times10^{15}cm^{-3}$.}}
	\label{fig3}
\end{figure}

In Fig.\ref{fig3}, we further demonstrate the influence of inter-valley scattering on the weight of intra-valley scattering. As shown in Fig.\ref{fig3}(a), the proportion of intra-valley scattering increases with the increase of $d$ value, and when the value of $d$ is large and the magnetic field is weak, intra-valley scattering dominates. In Fig.\ref{fig3}(b), it is noted that an increase in magnetic field strength leads to a significant increase in the proportion of inter-valley scattering, particularly in the weaker magnetic field range. The proportion of inter-valley scattering reaches a steady state as the magnetic field is increased up to a certain threshold(This corresponds to the case where $l_B$ is almost equal to $d$). In summary, at weak magnetic fields and small values of $d$, the intra-valley scattering process dominates. However, under strong magnetic fields and small $d$, the contributions of the inter-valley scattering process become significant(although the total transverse conductivity is small).
\begin{figure}[h]
	\centering
	\includegraphics[width=0.8\textwidth]{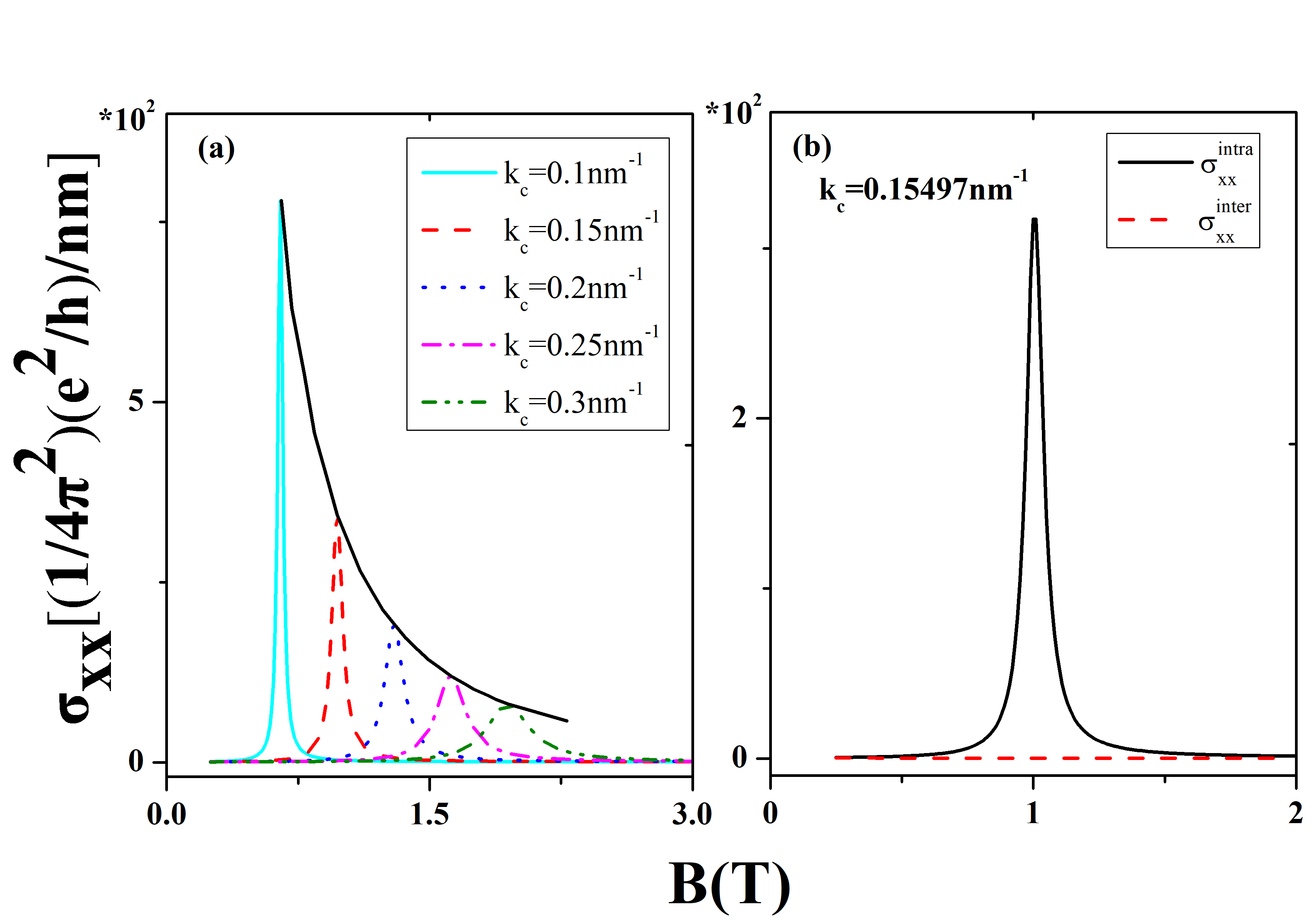}
	\caption{\small{The transverse magneto-conductivity, $\sigma_{xx}$, as a function of $B$ at different $k_c$(the positions of the Weyl nodes in the momentum space). In (a), the cyan solid line, the red dashed line, the blue dotted line, the pink dash-dot line and the green dash-dot-dot line are for $k_c=0.1nm^{-1}$, $k_c=0.15nm^{-1}$, $k_c=0.2nm^{-1}$, $k_c=0.25nm^{-1}$ and $k_c=0.3nm^{-1}$, respectively. And the black line is a curve synthesised from the resonance peak. In (b), the black solid line is for intra-valley scattering and the red dashed line is for inter-valley scattering, respectively. Here d=1.0nm, $n_{imp}=1.0\times10^{15}cm^{-3}$.}}
	\label{fig4}
\end{figure}
\subsection{Possible application of the measurement of the magnetoconductivity}
\label{sec33}
As we can see from subsections \ref{sec31} and \ref{sec32}, the magnetoconductivities induced by the charged impurities are determined by the magnetic field, the location of the Weyl nodes, and the distributions of the impurities. From an experimental perspective, the location of Weyl nodes in a undoped Weyl semimetal crystal sample with a known density of charged impurities may be determined by measuring magnetoconductivity. Conversely, the distribution of charged impurities can be deduced by measuring magnetoconductivity of a doped Weyl semimetal sample in which the Weyl node locations have already been determined. These two cases will be discussed below.

We first consider the case in a undoped Weyl semimetal crystal. As previously stated, the Fermi level shifts due to the introduction of charged impurities. To simulate the distribution of the impurity, we select the length scale $d=1nm$ in Eq.\eqref{eq26} to fix the Fermi level. The density of the impurity is fixed as $n_{imp}=1\times10^{15} cm^{-3}$. We adjust the value of $k_c$ and subsequently calculate $\sigma_{xx}$ based on this impurity distribution. The numerical results are shown in Fig.\ref{fig4}. From Fig.\ref{fig4}(a), it is evident that $\sigma_{xx}$ displays distinct peaks at different values of $k_c$ and $B$. As $k_c$ increases, or more precisely, as the distance between two Weyl nodes increases, the value of $\sigma_{xx}$ decreases. Furthermore, as $k_c$ increases, the peak position shifts to a higher value of $B$, and the change in $\sigma_{xx}$ becomes smoother. These results can be understood as follows: As the distance between the nodes increases, the probability of scattering decreases, resulting in a decrease in $\sigma_{xx}$. A stronger magnetic field is necessary to increase the probability of scattering and maintain magnetoconductivity.  From another perspective, the location of Weyl nodes can be obtained by fixing the value of $B$. As depicted in (b) of Fig.\ref{fig4}, if we set $B=1T$ , we can obtain the value of $k_c$, which is about $0.15497nm^{-1}$. This reminds us that we can use a smooth curve to fit the relationship between $k_c$ (i.e., the locations of the Weyl nodes) and $B$. Using non-linear fitting, we obtain the relations as $k_c=\frac{\alpha}{\left(B^{-\beta}\right)^{\gamma}}$, where $\alpha\approx0.15422, \beta\approx2.03193,\gamma\approx0.49202$. By measuring the magnetoconductivity, this equation may be used to determine the location of Weyl nodes in a Weyl semimetal sample with a known density of charged impurities.
\begin{figure}[h]
	\centering
	\includegraphics[width=0.8\textwidth]{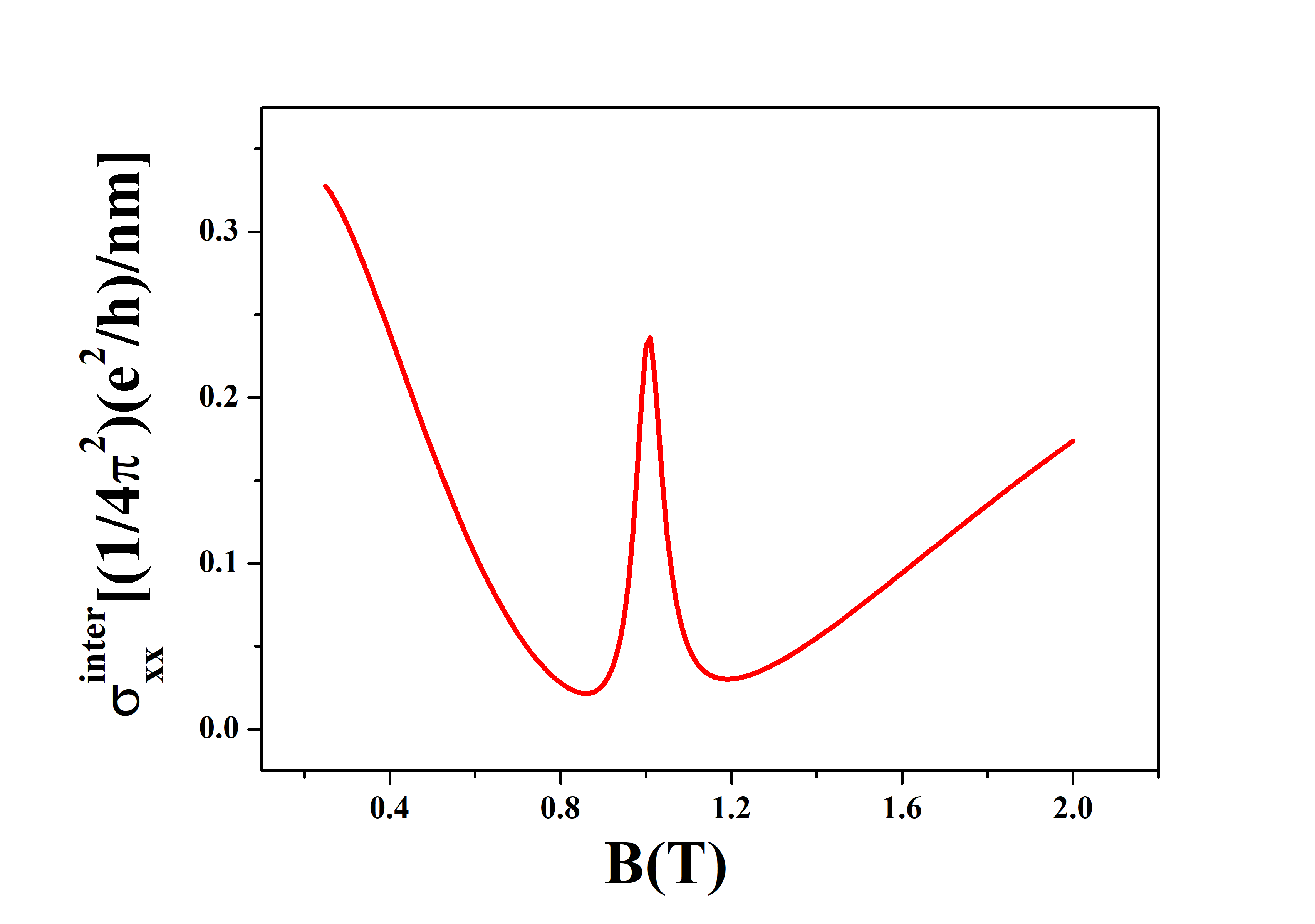}
	\caption{\small{The inter-valley scattering transverse magneto-conductivity, $\sigma_{xx}^{inter}$, as a function of $B$.  Here $d=1.0nm$, $k_c=0.15497nm$, $n_{imp}=1.0\times10^{15}cm^{-3}$.}}
	\label{fig5}
\end{figure}

As displayed in Fig.\ref{fig3}, the inter-valley scattering process has a significant impact on $\sigma_{xx}$ and should be considered. To verify this, using the same parameter selection as in Fig.\ref{fig4}(b), Fig.\ref{fig5} shows only the contributions of the inter-valley scattering process to $\sigma_{xx}^{inter}$. Compare with the intra-valley behavior $\sigma_{xx}^{intra}$ shown in Fig.\ref{fig4}(b), numerical results show that these two parts of conductivities have a quite different dependence on weak $B$. This can be understood as follows: a weak magnetic field corresponds to a relatively larger magnetic length $l_B$, which leads to $l_B$ being large enough to be compared with the typical scale of the impurity potential $d$. Under this circumstance, the electrons have a greater likelihood of participating in the inter-valley scattering process. When the magnetic field becomes strong, the $l_B$ decreases rapidly. This causes the electrons to become highly degenerated, ultimately leading to a decrease in conductivity. By calculating the first derivative of the expression for $\sigma_{xx}$ with respect to $B$, we confirmed this physical scenario. The numerical results indicate a negative first derivative of $\sigma_{xx}$ within the ranges of $B=0.2-0.8T$ and $B=1.0-1.16T$, and a positive first derivative within the ranges of $B=0.9-1.0T$ and $B=1.2-2.0T$.
\begin{figure}[h]
	\centering
	\includegraphics[width=0.8\textwidth]{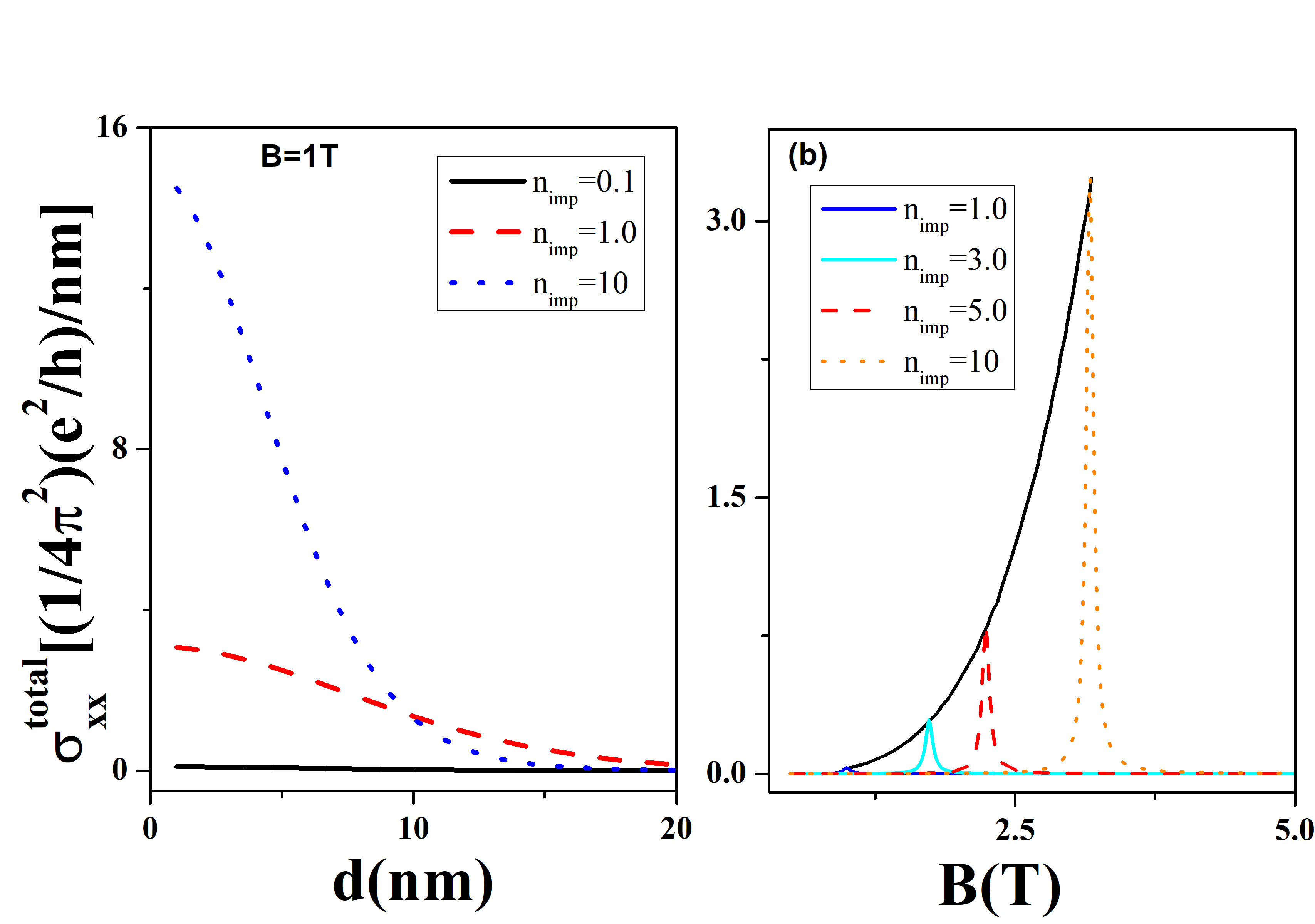}
	\caption{\small{The transverse magneto-conductivity, $\sigma_{xx,}$, as the functions of $d$ and $B$ at different $n_{imp}$(represent the doping degree). In (a), the black solid line is for $n_{imp}=0.1\times10^{15}cm^{-3}$, the red dashed line is for $n_{imp}=1.0\times10^{15}cm^{-3}$, and the blue dotted line is for $n_{imp}=0.1\times10^{15}cm^{-3}$, respectively. Here $B=1T$. In (b), the blue line is for $n_{imp}=0.1\times10^{15}cm^{-3}$, the cyan solid line is for $n_{imp}=3.0\times10^{15}cm^{-3}$, the red dashed line is for $n_{imp}=5.0\times10^{15}cm^{-3}$, and the orange dotted line is for $n_{imp}=10\times10^{15}cm^{-3}$, respectively. And the black solid line is a curve synthesized from the formant. Here $d=1.0nm$, $k_c=0.15497nm^{-1}$.}}
	\label{fig6}
\end{figure}
\par
Now we will compute the magnetoconductivity of a doped Weyl semimetal sample with incomplete information about the impurity distribution. Without loss of generality, we calculate the conductivity $\sigma_{xx}$ as a function of $d$ and $n_{imp}$ at a fixed magnetic field of $B=1T$ and $k_c=0.15497nm^{-1}$. Numerical results are shown in Fig.\ref{fig6}. From (a) we observe that $\sigma_{xx}$ increases with impurity density $n_{imp}$ and decreases with $d$. This reasoning is valid since a high density of impurities increases the probability of scattering, thereby resulting in a large $\sigma_{xx}$. On the other hand, a large value of $d$ corresponds to a widely-distributed impurity, which reduces the probability of scattering.
This scenario is based on intuition, however, in (a) we also observe a disproved outcome. As can be observed when $d$ approaches $10nm$, $\sigma_{xx}$ becomes smaller with $n_{imp}=10$ compared to the case with $n_{imp}=1$. We hypothesize that heavy doping increases the likelihood of electron collisions, thereby enhancing conductivity. However, excessive doping (also with a large $d$) inhibits electron collisions due to impurity “crowding”, thus causing a decrease in the probability of scattering and leading to a smaller $\sigma_{xx}$.

After comprehending the impact of impurity distribution on conductivity, our attention now turns to how to identify the impurity distribution through magnetoresistivity measurements. Similar as what we do in Fig.\ref{fig4}, we calculate the value of $\sigma_{xx}$ for various $n_{imp}$ while keeping $B=1T$, $d=1nm$, and $k_c=0.15497nm^{-1}$ fixed. Peaks also appears in $\sigma_{xx}$ and with a heavy doping, the peaks shifts to the right. Similarly, we fit a curve to show the relationship between the impurity density and the magnetic field: $n_{imp}=\eta B^{\lambda}$, where $\eta\approx1.00501$, $\lambda\approx0.5$. Similar, we may be able to use this property in the experiments to predict the doping level of the sample by measuring the magneticconductivity.
\begin{figure}[h]
	\centering
	\includegraphics[width=0.8\textwidth]{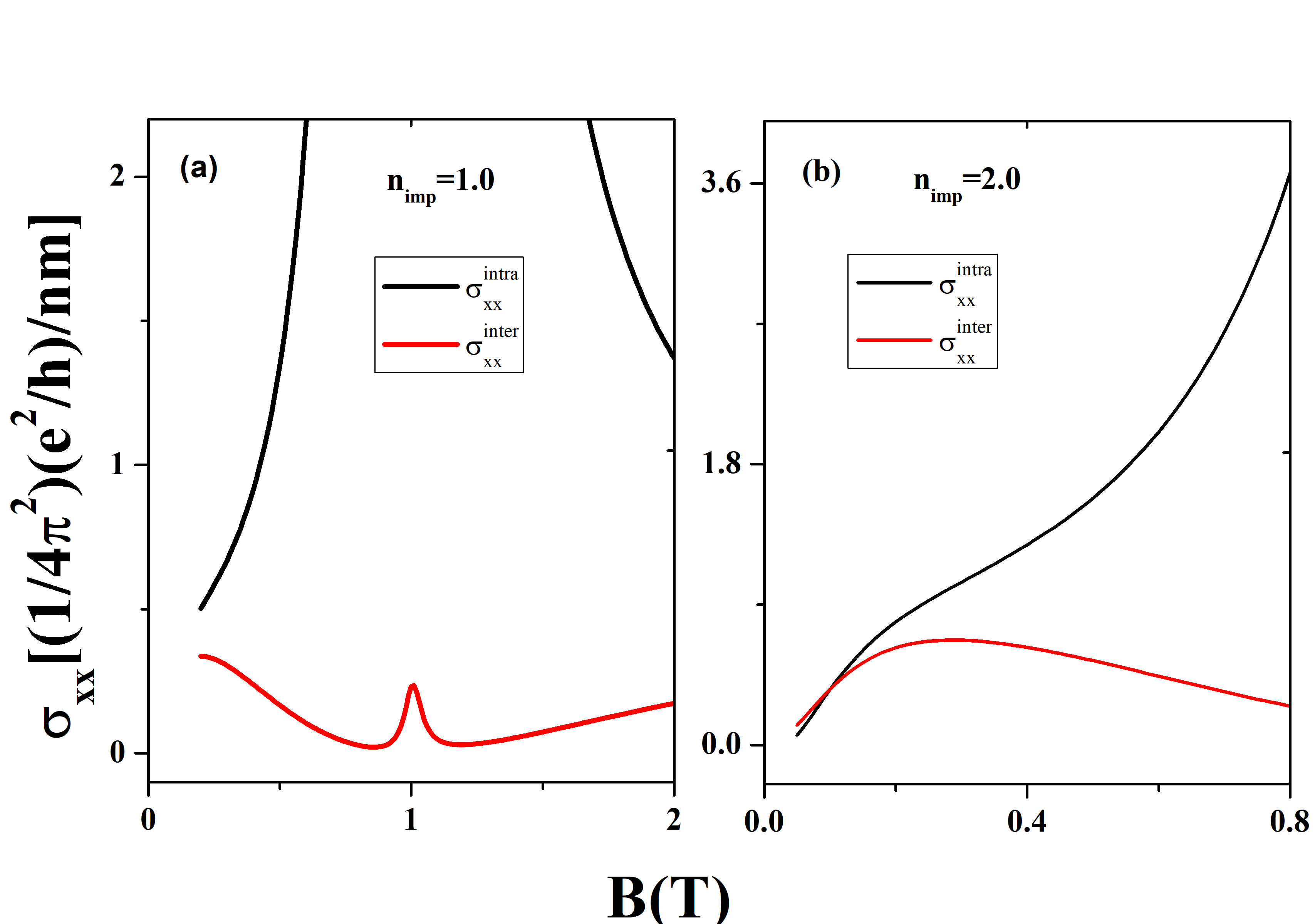}
	\caption{\small{The transverse magneto-conductivity, $\sigma_{xx}^{intra}$(the black line) and $\sigma_{xx}^{inter}$(the red line) , as function of $B$. The density of the impurity are $n_{imp}=1.0\times10^{15}cm^{-3}$in (a) and $n_{imp}=2.0\times10^{15}cm^{-3}$ in (b), respectively. Here $d=1.0nm$, $k_c=0.15497nm^{-1}$.}}
	\label{fig7}
\end{figure}

Finally, we continue check the contributions to the $\sigma_{xx}$ from the intra-valley and the inter-vally scattering process. We calculate $\sigma_{xx}$ as a function of $B$ with fixed value $d=1.0nm$ and $k_c=0.15497nm^{-1}$. The density of impurity are $n_{imp}=1.0\times10^{15}cm^{-3}$ in (a) and $n_{imp}=2.0\times10^{15}cm^{-3}$ in (b), respectively.
Similar as Fig.\ref{fig5}, there is a marked difference in the dependence of the intra-valley conductivity on the magnetic field strength compared to the inter-valley conductivity. From (a) in Fig.\ref{fig7}, we observe that at weak magnetic fields, the inter-valley conductivity closely approximates the intra-valley conductivity.\footnote{(a) in Fig.\ref{fig7} is just a part of (b) in Fig.\ref{fig4}.} When the density of the impurity increase to $n_{imp}=2.0\times10^{15}cm^{-3}$, it is observed that the inter-valley conductivity exceeds the intra-valley conductivity at $B=0.1T$, as shown in (b) in Fig.\ref{fig7}. These results provide additional evidence that electrons are more likely to engage in inter-valley scattering under weak magnetic field strengths, highlighting the considerable contributions of inter-valley scattering on magnetoconductivity.
\section{Conclusion and outlook}\label{sec4}
A Weyl semimetal sample found in nature usually contains various impurities, making it crucial to understand the structure of a specific sample for both theoretical and practical purposes. Measuring the magneto-conductivity resulting from impurities may aid in determining the distribution of impurities and revealing the structure of a Weyl semimetal sample. Based on this, we investigate the magneto-conductivity properties of a Weyl semimetal with two Weyl nodes. We employ a Gaussian random potential to model the impurity distribution and analyze the effect of charged impurities on both of the longitudinal and transverse magneto-conductivity.

By calculating the longitudinal magnetoc-onductivity, a definite positive correlation between magnetic flux density and the longitudinal magneto-conductivity is obtained. This suggests the presence of negative longitudinal magneto-resistivity observed in the experiments. Furthermore, the longitudinal magnetic conductivity is affected by the length scale parameter of the impurity distribution($d$). The study reveals that an increase in $d$ enhances the longitudinal magnetic conductivity, and when $d$ reaches a certain threshold(say $10 nm$ in Fig.\ref{fig1}), the longitudinal magnetic conductivity increases rapidly. These results may explain why negative magneto-resistivity has not been observed in experiments on each Weyl semimetal sample. Due to the varying distribution of impurities in a Weyl semimetal sample, certain distributions may result in a low longitudinal magnetic conductivity, which may become masked by other conductivities.

We then calculate the transverse magneto-conductivity by incorporating the contributions from both the intra-valley and inter-valley scattering processes. We find that the transverse magneto-conductivity is inversely proportional to the magnetic flux density. The inter-valley process becomes significant at a small distribution length scale of $d$ and a strong magnetic flux density of $B$. Another important aspect of transverse magneto-conductivity is the presence of peaks, depicted in Fig.\ref{fig4} and Fig.\ref{fig6}. Our calculations demonstrate that these peaks occur in proximity to the location of the Weyl nodes(identified as $k_c$). The results indicate that determining the position of Weyl nodes is possible by measuring the peak of transverse conductivity in combination with precise data on magnetic field and impurity density distribution. Alternatively, if the Weyl node location is fixed, one can deduce the distribution of impurities by measuring the peak of transverse conductivity with precise data on the applied magnetic field. Combined with the measured longitudinal magnetic conductivity data, one may gain insight into the structure of a natural sample of Weyl semimetal objectively and clearly.

As we know, multiple pairs of Weyl nodes exist in Weyl semimetal materials in the real world. In addition, it is possible to realize a state in which a Weyl semimetal coexists with a topological nodal line semimetal\upcite{Ji:2023rua}. Therefore, it is crucial to investigate the effect of impurities on transport in Weyl semimetals with multiple Weyl nodes or in a topological state where multiple topological phases coexist. Furthermore, the distribution of Weyl nodes and transport properties are also affected by strain. In future research, we will analyze the effects of strain on both one pair and multiple Weyl nodes, with the aim of providing a more accurate and realistic prediction of the structure of a Weyl semimetal sample.

\subsection*{Acknowledgments}
We would like to thank Zhengang Zhu and Zhi-Fan Zhang for useful discussions. This work is supported by National Natural Science Foundation of China (Grant No. 61974162).


\end{document}